\documentclass[twocolumn,prb,showpacs,aps]{revtex4-1}
\usepackage{amssymb}
\usepackage{amsmath}
\usepackage{graphicx}
\usepackage{epsfig}
\usepackage{xcolor}

\begin{document}

\title{Biot-Savart law in quantum matter}
\author{X. M. Yang and Z. Song}
\email{songtc@nankai.edu.cn}
\affiliation{School of Physics, Nankai University, Tianjin 300071,
China}

\begin{abstract}
We study the topological nature of a class of lattice systems, whose Bloch
vector can be expressed as the difference of two independent periodic vector
functions (knots) in an auxiliary space. We show exactly that each loop as a
degeneracy line generates a polarization field, obeying the Biot-Savart law:
The degeneracy line acts as a current-carrying wire, while the polarization
field corresponds to the generated magnetic field. Applying the Amp\`{e}re's
circuital law on a nontrivial topological system, we find that two Bloch
knots entangle with each other, forming a link with the linking number being
the value of Chern number of the energy band. In addition, two lattice
models, an extended QWZ model and a quasi-$1$D model with magnetic flux, are
proposed to exemplify the application of our approach. In the aid of the
Biot-Savart law, the pumping charge as a dynamic measure of Chern number is
obtained numerically from quasi-adiabatic processes.
\end{abstract}

\maketitle

\section{Introduction}

Condensed matter provides a platform to realize many physical objects in
other subjects such as Majorana and Dirac Weyl fermions which are proposed
in particle physics but not be discovered in nature \cite{XWAN, LLU, SMH,
HWENG, SYXU, BQLV, LLU2, VMO, SNA}. Another example is the Dirac monopole,
which is a point source of a magnetic field proposed by Dirac \cite{PAM}. It
has a quantum analogy in quantum physics, where the Berry curvature of
energy band acts as the magnetic field generated by degeneracy points as
Dirac monopoles \cite{DXIAO}.\ As the extension of degeneracy points, nodal
loops as closed $1$-dimensional ($1$D) manifolds in $3$D momentum space can
be classified as nodal rings \cite{AAB}, nodal chains \cite{TB}, nodal links
\cite{ZYAN}, and nodal knots \cite{RBI}. It has been extensively studied
both theoretically\cite{XQSUN, SNIE, JAHN, CFANG1, TK, JYL, CFANG2, PYC,
WCHEN, MEZAWA, YZHOU, ZYANG, MXIAO} and experimentally \cite{QXU, RYU, QYAN,
GCHANG, XFENG}. In the recent work, it has turned out that the relation
between degeneracy lines and the corresponding polarization field in the
parameter space is topologically isomorphic to Biot-Savart law in
electromagnetism \cite{RWANG}.

\begin{figure}[tbp]
\includegraphics[ bb=112 370 500 623, width=0.45\textwidth, clip]{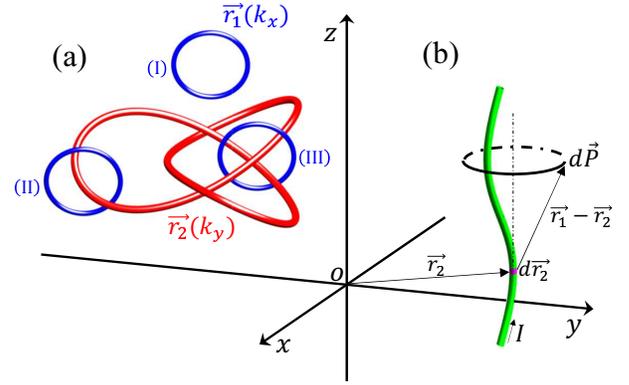}
\caption{Schematic illustration of the aim of present work. (a) We consider
a $2$D system with the Bloch Hamiltonian related to two periodic
vector functions $\mathbf{r}_{1}(k_{x})$
and $\mathbf{r}_{2}(k_{y})$ in auxiliary
space, which correspond to two knots. The topological index of the energy
band is determined by the relations of two knots: The Chern number of the
band equals to the linking number of two knots.\ Several representative
configurations of $[\mathbf{r}_{1}(k_{x}),$ $\mathbf{r}_{2}(k_{y})]$ is
presented. Here $\mathbf{r}_{2}$ is a trefoil knot, while three types of $%
\mathbf{r}_{1}$\ are taken as simple loops, but at different positions,
resulting linking numbers $\mathcal{N}=0$, $1$, and $2$, respectively. The
corresponding Bloch Hamiltonians describe the systems with Chern numbers $%
c=0 $, $1$, and $2$, respectively. (b) The main purpose of this work. For a $%
1$D model with a fixed function $\mathbf{r}_{2}(k_{y})$,  $\mathbf{r}_{2}(k_{y})$ is
referred as to the degeneracy circuit. For an
arbitrary point $\mathbf{r}_{1}(k_{x})$, the polarization filed $\mathbf{P}(%
\mathbf{r}_{1})$ obeys the Biot-Savart law for magnetic field arising from
the degeneracy loop as a current loop. Polarization field $\mathrm{d}\mathbf{%
P}$ at point $\mathbf{r}_{1}$\ generated from an infinitesmall length of
degeneracy line $\mathrm{d}\mathbf{r}_{2}$ at $\mathbf{r}_{2}$, has the
identical form with the Biot-Savart Law related magnetic field generated
from the current loop. Finding the polarization field $\mathbf{P}$ at
arbitrary point $\mathbf{r}_{1} $ resulting from a degeneracy line can be
simply obtained by the Biot-Savart law in electromagnetism.}
\label{fig1}
\end{figure}

In this work, we provide another quantum analogy of classical
electromagnetism. We consider a class of Bloch Hamiltonians, which contains
two periodic vector functions with respect to two independent variables,
such as momentum $k_{x}$\textbf{\ }and\textbf{\ }$k_{y}$ for a $2$D lattice
system, respectively. These two periodic vector functions correspond to two
knots in $3$D auxiliary space (see Fig. \ref{fig1}(a)). The Bloch vector is
the difference of two vectors. When we only consider one of two knots, the
system reduces to a $1$D lattice system. The Zak phase and polarization
field at a fixed point in $3$D auxiliary space can be obtained.\ We show
exactly that the knot as a degeneracy line has a simple relation of its
corresponding polarization field, obeying the Biot-Savart law: The
degeneracy line acts as a current-carrying wire, while the polarization
field corresponds to the generated magnetic field. The relationship between
two knots can be characterized by applying the Amp\`{e}re's circuital law on
the field integral arising from one knot along another knot. For a
nontrivial topological system, the integral is nonzero, due to the fact that
two Bloch knots entangle with each other, forming a link with the linking
number being the value of Chern number of the energy band. In Fig. \ref{fig1}%
, we schematically illustrate the main conclusion of this work.

We propose two lattice models to exemplify the application of our approach.
The first one is an extended QWZ model. We show that the Bloch Hamiltonian
is an example of our concerned system. Two knots of the original QWZ model
simply reduce to two circles. The second one is a time-dependent quasi-$1$D
model with magnetic flux. In this case, the Amp\`{e}re circulation integral
is equivalent to the topological invariant. In the aid of the Biot-Savart
law, the pumping charge acts as a dynamic measure of the Chern number. We
perform numerical simulation for several representative quasi-adiabatic
processes to demonstrate this application.

The remainder of this paper is organized as follows. In Sec. \ref{Model with
degeneracy loop}, we present a class of models, whose Bloch Hamiltonian
relates to two knots. In Sec. \ref{Kitaev model on square lattice} We
propose the extended QWZ model to exemplify the application of our approach.
Sec. \ref{Ladder system} gives another example, which is a time-dependent
quasi-$1$D model with magnetic flux. Sec. \ref{Pumping charge} devotes to a
dynamic measure of Chern number, the pumping charge, which can be computed
numerically for several representative quasi-adiabatic processes to
demonstrate our work. Finally, we present a summary and discussion in Sec. %
\ref{Summary}.

\section{Double-knot model}

\label{Model with degeneracy loop}

Consider a Bloch Hamiltonian $h_{\mathbf{k}}$ in the form%
\begin{eqnarray}
h_{\mathbf{k}} &=&\left(
\begin{array}{cc}
\left( z_{1}-z_{2}\right) & x_{1}-x_{2}-i\left( y_{1}-y_{2}\right) \\
x_{1}-x_{2}+i\left( y_{1}-y_{2}\right) & -\left( z_{1}-z_{2}\right)%
\end{array}%
\right)  \notag \\
&=&\left[ \mathbf{r}_{1}(k_{x})\mathbf{-r}_{2}(k_{y})\right] \mathbf{\cdot
\sigma },  \label{hk}
\end{eqnarray}%
which is the starting point of our study. It is consisted of two periodic
vector functions $\mathbf{r}_{1}(k_{x})=\mathbf{r}_{1}(2\pi +k_{x})=x_{1}%
\mathbf{i}+y_{1}\mathbf{j}+z_{1}\mathbf{k}$\textbf{\ }and\textbf{\ }$\mathbf{%
r}_{2}(k_{y})=\mathbf{r}_{2}(2\pi +k_{y})=x_{2}\mathbf{i}+y_{2}\mathbf{j}%
+z_{2}\mathbf{k}$, representing two knots (loops) in $3$D auxiliary space.
Here $\mathbf{\sigma =(}\sigma _{x},\sigma _{y},\sigma _{z}\mathbf{)}$ are
Pauli matrices and $h_{\mathbf{k}}$\ represents a class of models, which is
referred\ as to double-knot (double-loop) model. Matrix $h_{\mathbf{k}}$\
can take the role of a core matrix of crystalline system for non-interacting
Hamiltonian, or Kitaev Hamiltonian. We note that the spectrum of $h_{\mathbf{%
k}}$\ is two-band and the gap closes when two knots have crossing points.
The aim of this work is to reveal the feature of the system which is
originated from the character of two knots.

To this end, we first consider the case with a fixed $k_{x}$. Then the model
only contains a point $\mathbf{r}_{1}$ and a knot $\mathbf{r}_{2}\left(
k_{y}\right) $. The Hamiltonian reduces to%
\begin{equation}
h_{k_{y}}=\left[ \mathbf{r}_{1}\mathbf{-r}_{2}(k_{y})\right] \mathbf{\cdot
\sigma },
\end{equation}%
which is a $1$D system in real space. Here $r_{2}(k_{y})$\ is a degeneracy
line, at which the gap closes.\ The solution of equation $%
h_{k_{y}}\left\vert u_{\pm }^{k_{y}}\right\rangle =\varepsilon _{k_{y}}^{\pm
}\left\vert u_{\pm }^{k_{y}}\right\rangle $ has the form
\begin{equation}
\left\vert u_{+}^{k_{y}}\right\rangle =\left(
\begin{array}{c}
\cos \frac{\theta _{k_{y}}}{2}e^{-i\varphi _{k_{y}}} \\
\sin \frac{\theta _{k_{y}}}{2}%
\end{array}%
\right) ,\left\vert u_{-}^{k_{y}}\right\rangle =i\left(
\begin{array}{c}
-\sin \frac{\theta _{k_{y}}}{2} \\
\cos \frac{\theta _{k_{y}}}{2}e^{i\varphi _{k_{y}}}%
\end{array}%
\right)
\end{equation}%
with $\varepsilon _{k_{y}}^{\pm }=\pm \left\vert \mathbf{r}_{1}\mathbf{-r}%
_{2}\left( k_{y}\right) \right\vert $, where the azimuthal and polar angles
are defined as%
\begin{equation}
\cos \theta _{k_{y}}=\frac{z_{1}-z_{2}}{\left\vert \mathbf{r}_{1}\mathbf{-r}%
_{2}\right\vert },\tan \varphi _{k_{y}}=\frac{y_{1}-y_{2}}{x_{1}-x_{2}}.
\end{equation}%
For this $1$D system, the corresponding Zak phases for upper and lower bands
are defined as%
\begin{equation}
\mathcal{Z}_{\pm }=\frac{i}{2\pi }\int_{-\pi }^{\pi }\left\langle u_{\pm
}^{k_{y}}\right\vert \frac{\partial }{\partial k_{y}}\left\vert u_{\pm
}^{k_{y}}\right\rangle \mathrm{d}k_{y}.
\end{equation}%
It is well known that the Zak phase is gauge-dependent and the present
expression of $\left\vert u_{\pm }^{k_{y}}\right\rangle $\ results in%
\begin{equation}
\mathcal{Z}=\mathcal{Z}_{+}=-\mathcal{Z}_{-}=\frac{1}{2\pi }\oint_{\mathrm{L}%
}\cos ^{2}\frac{\theta _{k_{y}}}{2}\mathrm{d}\varphi _{k_{y}},
\end{equation}%
where \textrm{L} denotes the integral loop about the solid angle.
Accordingly, the polarization vector field is defined as%
\begin{equation}
\mathbf{P}=-\mathbf{\nabla }\mathcal{Z},
\end{equation}%
where $\mathbf{\nabla }$ is the nabla operator%
\begin{equation}
\mathbf{\nabla }=(\frac{\partial }{\partial x_{1}}\mathbf{i}+\frac{\partial
}{\partial y_{1}}\mathbf{j}+\frac{\partial }{\partial z_{1}}\mathbf{k}),
\end{equation}%
with unitary vectors $\mathbf{i}$, $\mathbf{j}$, and $\mathbf{k}$ in $3$D
auxiliary space. Straightforward derivation (see Appendix) shows that

\begin{equation}
\mathbf{P}=\frac{1}{4\pi }\oint_{\mathrm{L}}\frac{\mathrm{d}\mathbf{r}%
_{2}\times \left( \mathbf{r}_{1}-\mathbf{r}_{2}\right) }{\left\vert \mathbf{r%
}_{1}-\mathbf{r}_{2}\right\vert ^{3}},  \label{Polarization}
\end{equation}%
where $\mathrm{L}$ denotes the integral loop about the degeneracy loop. It
is clear that if we consider a degeneracy loop as current-carrying wire with
steady current strength $I=1/\mu _{0}$, flowing in the direction of
increasing $k_{y}$ from $0$\ to $2\pi $, the field $\mathbf{P}$ is identical
to the magnetic field generated by the current loop,\ where $\mu _{0}$\ is
the vacuum permittivity of free space. Since the Eq. (\ref{Polarization})
holds for an arbitrary loop $\mathrm{L}$, one can have its differential form%
\begin{equation}
\mathrm{d}\mathbf{P}=\frac{1}{4\pi }\frac{\mathrm{d}\mathbf{r}_{2}\times
\left( \mathbf{r}_{1}-\mathbf{r}_{2}\right) }{\left\vert \mathbf{r}_{1}-%
\mathbf{r}_{2}\right\vert ^{3}},  \label{Biot-Savart law}
\end{equation}%
which is illustrated in Fig. \ref{fig1}. It indicates that the relationship
between $\mathbf{P}$ and the degeneracy loop obeys the Biot-Savart law. It
reveals the topological characteristics of the degeneracy lines in a clear
physical picture. We will regard degeneracy loops as a \textit{band
degeneracy circuit}. This result helps us to determine the polarization of
any loops in the auxiliary space. In addition, the Amp\`{e}re circulation
integral $\oint_{\ell }\mathbf{P(\mathbf{r})\cdot }\mathrm{d}\mathbf{r}$
along a loop $\ell $\ has clear physical means: (i) It equals to the sum of
the current through the surface spanned by the loop $\ell $; (ii) It is the
pumping charge for the adiabatic passage $\ell $.

\begin{figure*}[tbp]
\includegraphics[ bb=3 414 620 765, width=0.92\textwidth, clip]{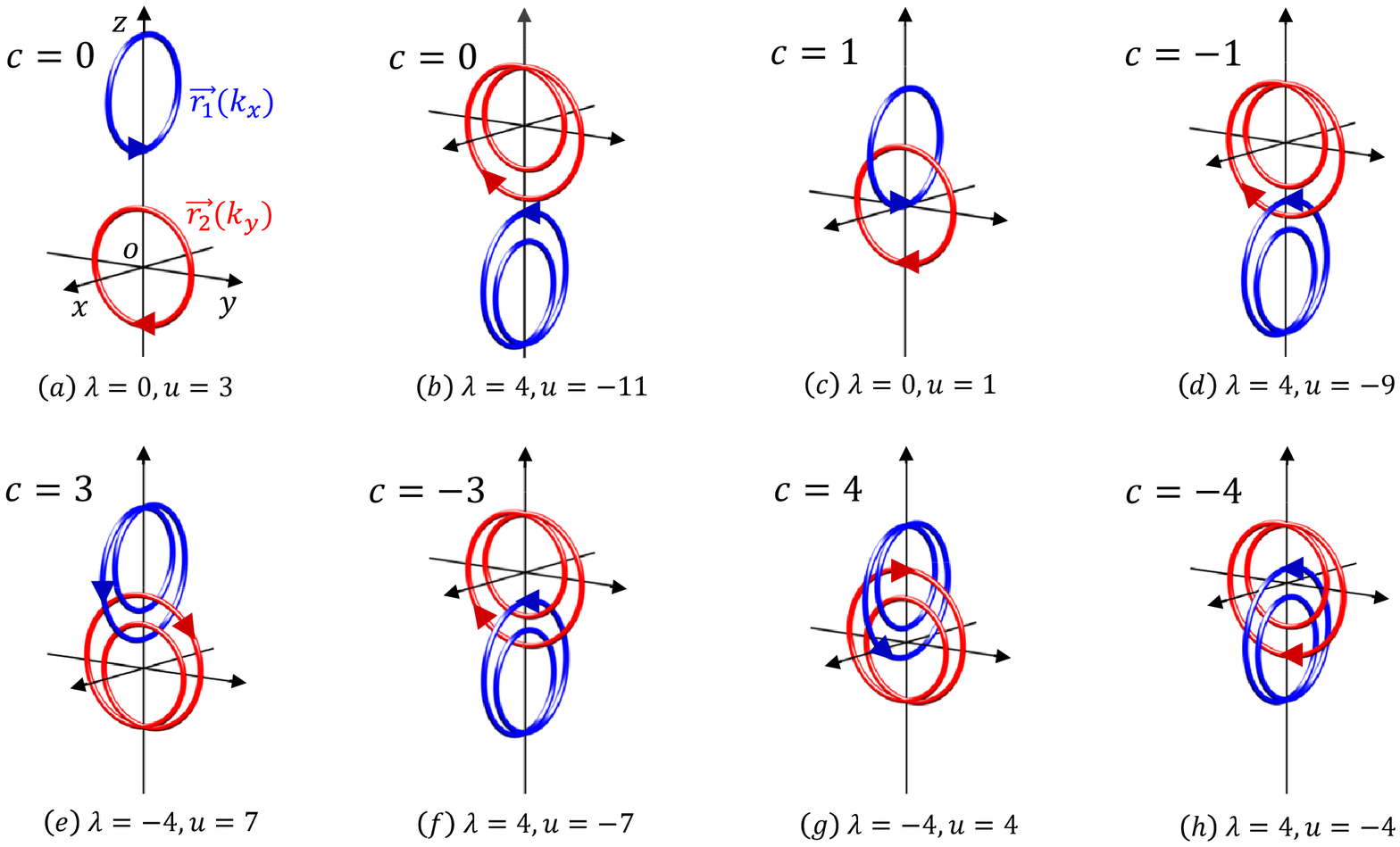}
\caption{Schematic several representative configurations of double-knot $%
\left\{ \mathbf{r}_{1}(k_{x}),\mathbf{r}_{2}(k_{y})\right\} $ for the
extended QWZ model. The plots are obtained from parameter equations in Eq. (%
\protect\ref{XYZ}) with parameters indicated in the panels. The arrows on
the loops indicate the directions of the knots with various topologies. The
corresponding Chern numbers are labeled, that match the linking numbers
exactly.}
\label{fig2}
\end{figure*}

Now we go back to $h_{\mathbf{k}}$, taking the loop $\ell $\ as the knot $%
\mathbf{r}_{1}(k_{x})$, which has no crossing point on the knot $\mathbf{r}%
_{2}(k_{y})$. We find that the corresponding Amp\`{e}re circulation integral
is connected to the topology of two knots and the band structure of the
system
\begin{equation}
-\oint_{\ell }\mathbf{P(\mathbf{r})\cdot }\mathrm{d}\mathbf{r=}c=\mathcal{N}%
\mathbf{.}  \label{CN}
\end{equation}%
Here the Chern number for lower band is defined as \cite{XLQI, GYCHO}%
\begin{equation}
c=\frac{1}{4\pi }\int_{0}^{2\pi }\int_{0}^{2\pi }\frac{\mathbf{r}^{\prime }}{%
\left\vert \mathbf{r}^{\prime }\right\vert ^{3}}\mathbf{\cdot }\left( \frac{%
\partial \mathbf{r}^{\prime }}{\partial k_{x}}\times \frac{\partial \mathbf{r%
}^{\prime }}{\partial k_{y}}\right) \mathrm{d}k_{x}\mathrm{d}k_{y},
\end{equation}%
with $\mathbf{r}^{\prime }=\mathbf{r}_{1}-\mathbf{r}_{2}$, which also equals
to the linking number \cite{RICCA} of two knots $\mathbf{r}_{1}(k_{y})$ and $%
\mathbf{r}_{2}(k_{x})$
\begin{equation}
\mathcal{N}=\frac{1}{4\pi }\int_{0}^{2\pi }\int_{0}^{2\pi }\frac{\mathbf{r}%
^{\prime }}{\left\vert \mathbf{r}^{\prime }\right\vert ^{3}}\mathbf{\cdot }%
\left( \frac{\partial \mathbf{r}_{1}}{\partial k_{x}}\times \frac{\partial
\mathbf{r}_{2}}{\partial k_{y}}\right) \mathrm{d}k_{x}\mathrm{d}k_{y}.
\end{equation}%
These relations are evident demonstrations of the system's topological
feature and clearly reveal the physical significance of the Amp\`{e}re
circulation integral $\oint_{\ell }\mathbf{P(\mathbf{r})\cdot }\mathrm{d}%
\mathbf{r}$. Furthermore, it corresponds to the jump of Zak phase for an
adiabatic passage along a knot, which\ can be measured by the Thouless
pumping charge in a quasi $1$D system. In the following, we present two
examples to illustrate our results.

\section{Extended QWZ model}

\label{Kitaev model on square lattice}

In this section, we consider a model, which is an extension of QWZ model
introduced by Qi, Wu and Zhang \cite{QWZ}, to illustrate our result. The
Bloch Hamiltonian is

\begin{equation}
h_{\mathbf{k}}=B_{x}\sigma _{x}+B_{y}\sigma _{y}+B_{z}\sigma _{z},
\end{equation}%
where the field components are%
\begin{equation}
\left\{
\begin{array}{l}
B_{x}=\sin k_{x}+\lambda \sin \left( 2k_{x}\right) \\
B_{y}=\sin k_{y}+\lambda \sin \left( 2k_{y}\right) \\
B_{z}=u+\cos k_{x}+\cos k_{y} \\
+\lambda \left[ \cos \left( 2k_{x}\right) +\cos \left( 2k_{y}\right) \right]%
\end{array}%
\right. .  \label{XYZ}
\end{equation}%
It reduces to original QWZ model when taking $\lambda =0$.

Now we rewrite it in the form%
\begin{equation}
h_{\mathbf{k}}=\left[ \mathbf{r}_{1}(k_{x})\mathbf{-r}_{2}(k_{y})\right]
\mathbf{\cdot \sigma },
\end{equation}%
where two vector functions are%
\begin{equation}
\left\{
\begin{array}{l}
\mathbf{r}_{1}=(\sin k_{x}+\lambda \sin \left( 2k_{x}\right) ,0,u+\cos
k_{x}+\lambda \cos \left( 2k_{x}\right) ) \\
\mathbf{r}_{2}=-(0,\sin k_{y}+\lambda \sin \left( 2k_{y}\right) ,\cos
k_{y}+\lambda \cos \left( 2k_{y}\right) )%
\end{array}%
\right. .
\end{equation}%
It is clear that $\mathbf{r}_{1}(k_{x})$\ and $\mathbf{r}_{2}(k_{y})$\
represent two limacons within $xz$ and $yz$ plane, respectively. When taking
$\left\vert \lambda \right\vert <0.5$, the crossing point of the limacon
disappears. Particularly, when taking $\lambda =0$, limacons reduce to
circles. The radiuses of two circles are both $1$, but the centers are $%
(0,0,u)$\ and $(0,0,0)$, respectively. Chern numbers can be easily obtained
from the linking numbers of these two circles: $c=0$, for $\left\vert
u\right\vert >2$, and $c=\pm 1$, for $0<\pm u<2$. When taking $\left\vert
\lambda \right\vert >0.5$, the crossing point of the limacon appears. Since
limacons with crossing point cannot be classified as knots, we add
perturbation terms $\kappa \sin 2k_{x}$\ to $r_{1y}$ and $\kappa \sin 2k_{y}$
to\ $r_{2x}$\ to untie the crossing point ($\left\vert \kappa \right\vert
\ll 1$), then limacons become knots again. The possible linking numbers of
such two knots are still equal to the Chern numbers $c=0$, $\pm 1$, $\pm 3$,
and $\pm 4$. The absence of $c=\pm 2$ is due to the fact that we take the
identical $\lambda $ in the expressions of $\mathbf{r}_{1}(k_{x})$ and $%
\mathbf{r}_{2}(k_{y})$. In Fig. \ref{fig2}, we plot some representative
configurations to demonstrate this point. Comparing to the direct
calculation of Chern number from the Berry connection, the example shows
that the Chern number can be easily obtained by the geometrical
configurations hidden in the Bloch Hamiltonian.

\section{Ladder system}

\label{Ladder system}

As a simple application of our result, we consider a quasi $1$D system with
periodically time-dependent parameters. The Bloch Hamiltonian has the form%
\begin{equation}
h_{k}(t)=\left[ \mathbf{r}(t)\mathbf{-r}_{c}(k)\right] \mathbf{\cdot \sigma }%
,
\end{equation}%
where $\mathbf{r}(t)=\mathbf{r}(t+T)$ represents a loop $\ell $ without
crossing point on the degeneracy loop $\mathbf{r}_{c}(k)$. The result
obtained above still apply to the case of replacing $(k_{x},k_{y})$\ with $%
(t,k)$, and replacing $\left\vert u_{\pm }^{\mathbf{k}}\right\rangle $ with $%
|u_{\pm }^{k}(t)\rangle $ accordingly. In this section we will demonstrate
our result and its physical implications through an alternative
tight-binding model, which is two coupled SSH chains, or a ladder system
with staggered magnetic flux, on-site potential and long range hopping
terms. These ingredients allow the system to support multiple types of
degeneracy loops with different geometric topologies.

\begin{figure}[tbp]
\includegraphics[ bb=130 440 476 540, width=0.5\textwidth, clip]{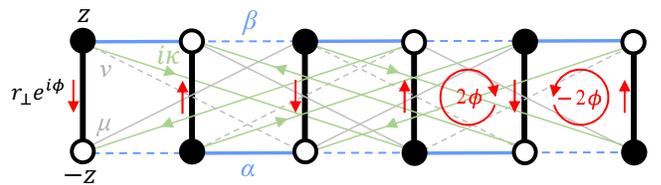}
\caption{Schematics of the two coupled SSH chains with staggered flux and
potential. The system consists of two sublattices A and B with on-site
potentials $z$\ and $-z$, indicated by filled and empty circles,
respectively. Hopping amplitudes along each chain are staggered by $\protect%
\alpha $ (blue solid line) and $\protect\beta $ (blue dotted line). The
interchain hopping amplitude is $r_{\perp }$ (thick black line) associated
with a phase factor and interchain diagonal hopping amplitude $\protect\mu $
(gray solid line), $\protect\nu $ (gray dotted line) and $i\protect\kappa $
(light green solid line). The red arrows indicate the hopping directions for
complex amplitudes, which are induced by the staggered flux threading each
plaquettes (arrow circles).}
\label{fig3}
\end{figure}

\begin{figure*}[tbp]
\includegraphics[ bb=19 8 1241 380, width=0.92\textwidth, clip]{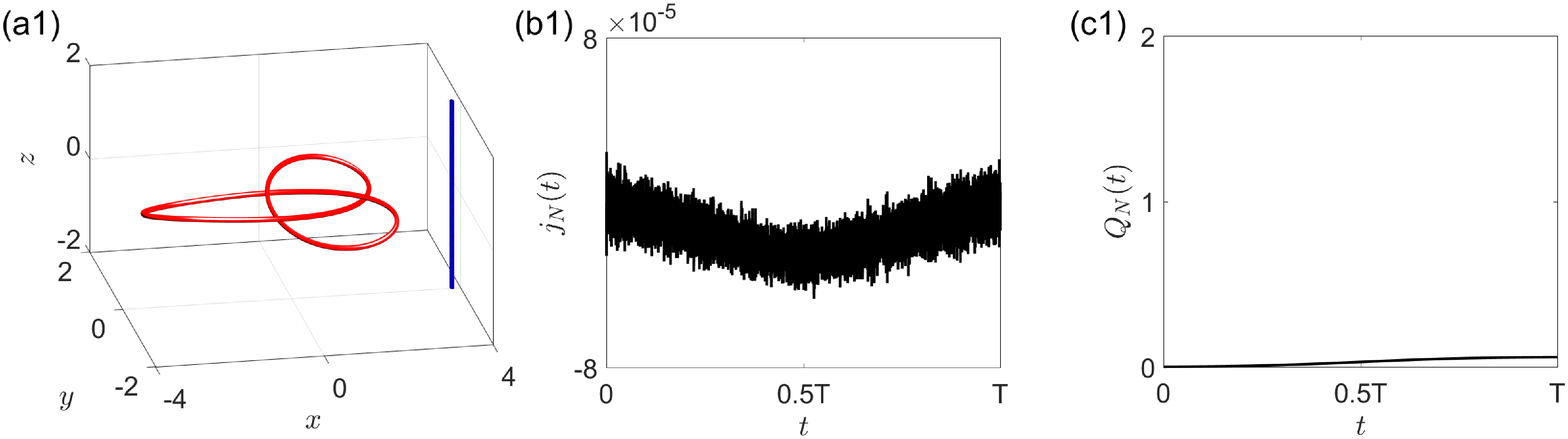} %
\includegraphics[ bb=19 8 1241 380, width=0.92\textwidth, clip]{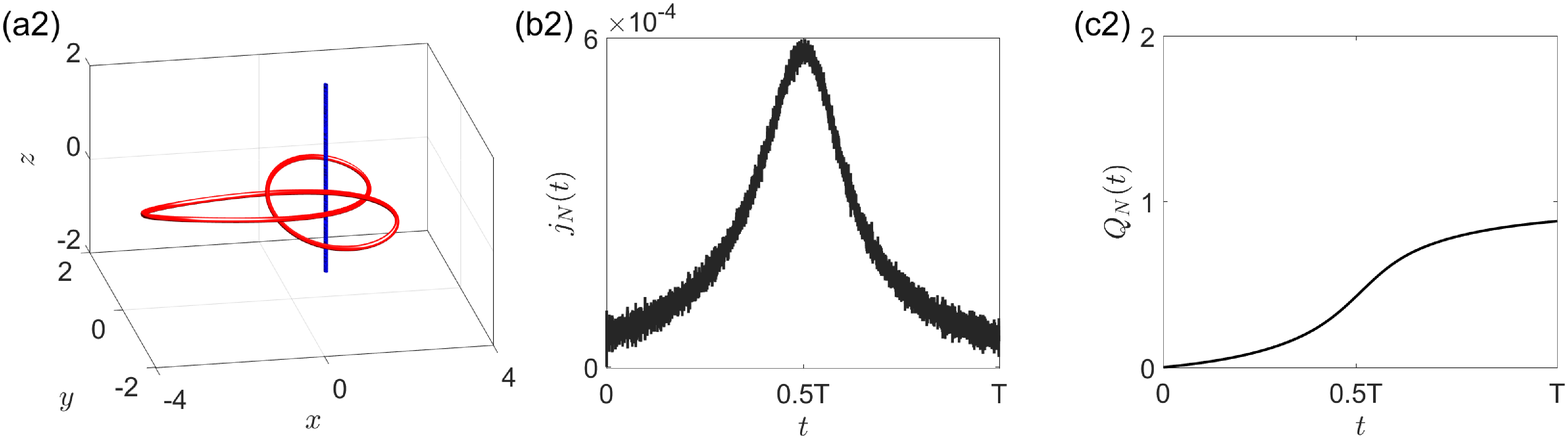} %
\includegraphics[ bb=19 8 1241 380, width=0.92\textwidth, clip]{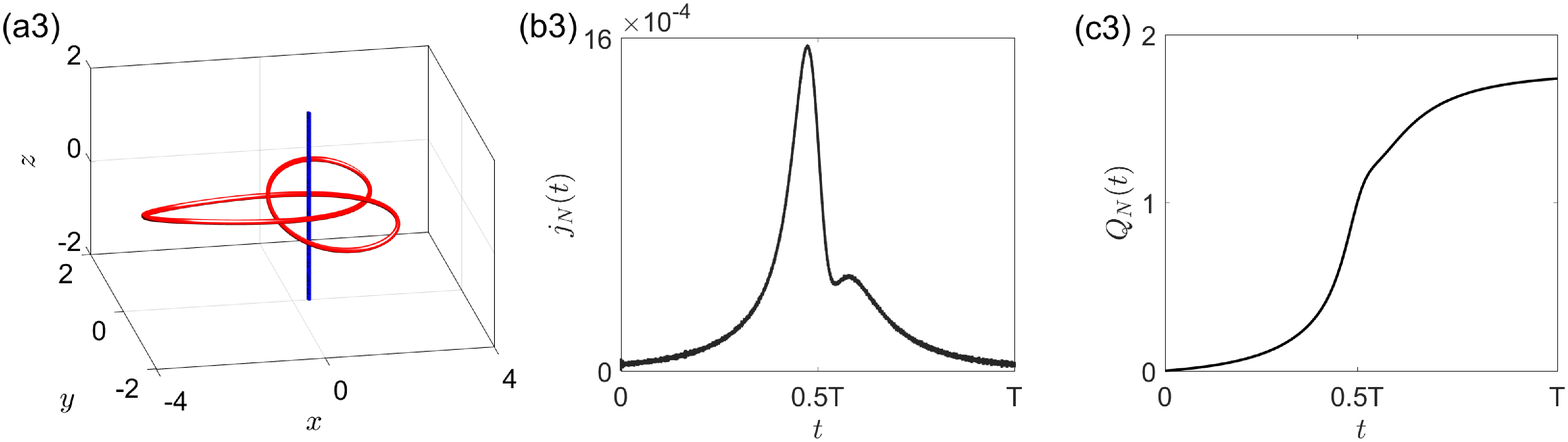}
\caption{ (a1-a3) Schematics of three adiabatic passages in $3$D auxiliary
space for pumping charge. The degeneracy curve (red) is a trefoil knot with
parameters $\protect\alpha =\protect\mu =0.5$, $\protect\beta =1$, $\protect%
\nu =1.5$, and $\protect\kappa =0.1$ of the system in Eq. (\protect\ref{ladd}%
) (Fig. \protect\ref{fig3}). The adiabatic passages are straight lines
(blue)\ at positions $(x,y):$ (a1) $(3.80,0)$, (a2) $(1.15,0.84)$, and (a3) $%
(0.40,0.01)$, respectively. (b1-b3) and (c1-c3)\ are plots of current and
the corresponding total charge transfer for quasi-adiabatic process. The
results are obtained by numerically exact diagonalization method for the
system in Eq. (\protect\ref{ladd}) with $N=100 $. The speed of time
evolution is $\protect\omega =1\times 10^{-3}$. It indicates that the
topological invariant can be obtained by dynamical process.}
\label{fig4}
\end{figure*}

We consider a ladder system which is illustrated in Fig. \ref{fig3},
represented by the Hamiltonian

\begin{eqnarray}
&&H_{\text{L}}=\sum_{j=1}^{N}\{r_{\bot }e^{i\phi }c_{2j}^{\dag
}c_{2j-1}+\alpha c_{2j-1}^{\dag }c_{2\left( j+1\right) }+\beta c_{2\left(
j+1\right) -1}^{\dag }c_{2j}  \notag \\
&&+\mu c_{2\left( j+2\right) -1}^{\dag }c_{2j}+\nu c_{2j-1}^{\dag
}c_{2\left( j+2\right) }+i\kappa \lbrack c_{2\left( j+3\right) -1}^{\dag
}c_{2j-1}  \notag \\
&&-c_{2\left( j+3\right) }^{\dag }c_{2j}]+\text{\textrm{H.c.}}%
\}+z\sum_{j=1}^{2N}\left( -1\right) ^{j+1}c_{j}^{\dag }c_{j},\label{ladd}
\end{eqnarray}%
on a $2N$ lattice. Here $c_{j}^{\dag }$ is the creation operator of a
fermion at the $j$th site with the periodic boundary condition $%
c_{2N+1}=c_{1}$. The inter-sublattice hopping amplitudes are $\left( \alpha
,\beta ,\mu ,\nu \right) $ and the intra-sublattice hopping amplitude is $%
\kappa $.\ Besides, two time-dependent parameters, $2\phi (t)$ is the
staggered magnetic flux threading each plaquette and $z(t)$\ is the strength
of staggered potentials. The ladder system is essentially two coupled SSH
chains. As a building block of the system, the SSH model \cite{SSH} has
served as a paradigmatic example of the $1$D system supporting topological
character \cite{Zak}. It has an extremely simple form but well manifests the
typical feature of topological insulating phase, and the transition between
non-trivial and trivial topological phases, associated with the number of
zero energy and edge states as the topological invariant \cite{Asboth}. It
has been demonstrated that all the parameters of this model can be easily
accessed within the existing technology of cold-atomic experiments \cite%
{Clay,Ueda,Jo}. We schematically illustrate this model in Fig.~\ref{fig3}.
We introduce the fermionic operators in $k$ space
\begin{equation}
\left\{
\begin{array}{l}
a_{k}=\frac{1}{\sqrt{N}}\sum_{j=1}^{N}e^{-ikj}c_{2j-1} \\
b_{k}=\frac{1}{\sqrt{N}}\sum_{j=1}^{N}e^{-ikj}c_{2j}%
\end{array}%
\right. ,  \label{Fourier 1}
\end{equation}%
and the wave vector $k=\pi (2n-N)/N$, $(n=0,1,...,N-1)$. Then we have%
\begin{equation}
H_{\text{L}}=\sum_{k}(a_{k}^{\dagger },b_{k}^{\dagger })h_{k}\left(
\begin{array}{c}
a_{k} \\
b_{k}%
\end{array}%
\right) ,
\end{equation}%
where the core matrix has the form%
\begin{equation}
h_{k}=\left(
\begin{array}{cc}
z+2\kappa \sin \left( 3k\right) & R(\phi ,k) \\
R^{\ast }(\phi ,k) & -z-2\kappa \sin \left( 3k\right)%
\end{array}%
\right) ,
\end{equation}%
and the off-diagonal matrix element is%
\begin{equation}
R(\phi ,k)=r_{\bot }e^{-i\phi }+\alpha e^{ik}+\beta e^{-ik}+\mu e^{-2ik}+\nu
e^{2ik}.
\end{equation}%
Taking

\begin{equation}
x+iy=r_{\bot }e^{i\phi },
\end{equation}%
the parameter equations for degeneracy loop is%
\begin{equation}
\left\{
\begin{array}{l}
x_{c}=-\left( \alpha +\beta \right) \cos k-\left( \mu +\nu \right) \cos
\left( 2k\right) \\
y_{c}=-\left( \beta -\alpha \right) \sin k-\left( \mu -\nu \right) \sin
\left( 2k\right) \\
z_{c}=-2\kappa \sin \left( 3k\right)%
\end{array}%
\right. ,  \label{trefoil knot}
\end{equation}%
which is plotted in Fig. \ref{fig4} for the case with parameters $\alpha
=\mu =0.5$, $\beta =1$, $\nu =1.5$, and $\kappa =0.1$. One can see that the
degeneracy curve is a trefoil knot. Intuitively, it should result in
topological features with indices $2$, $1$, and $0$. We will demonstrate
this point in the next section.

\section{Pumping charge}

\label{Pumping charge}

For a $2$D system with Bloch Hamiltonian in the form of Eq. (\ref{hk}), the
physical and geometric meanings of\ Chern number is well established. For a
quasi $1$D system with Bloch Hamiltonian in the form of Eq. (\ref{hk}) by
replacing $(k_{x},k_{y})$\ with $(t,k)$, the Chern number is connected to an
adiabatic passage driven by the parameters from $t$\ to $t+T$, or a periodic
loop $\mathbf{r=r}(t)$ in auxiliary space. In a $1$D model, it has been
shown that the adiabatic particle transport over a time period takes the
form of the Chern number\ and it is quantized \cite{DXIAO}. The pumped
charge counts the net number of degeneracy point enclosed by the loop. This
can be extended to the loop $\mathbf{r=r}(t)$\ in the present model.

Actually, one can rewrite Eq. (\ref{CN}) in the form%
\begin{equation}
c=\mathcal{N}=-\oint_{\ell }\mathbf{P(\mathbf{r})\cdot }\frac{\partial
\mathbf{\mathbf{r}}}{\partial t}\mathrm{d}t\mathbf{.}
\end{equation}%
where $\mathbf{\mathbf{r}}$\ (or $r_{\bot },\phi ,$ and $z$) is periodic
function of time $t$. Furthermore, we can find out the physical meaning of
the Chern number by the relation
\begin{equation}
c=\int_{0}^{T}\mathcal{J}(t)\mathrm{d}t,
\end{equation}%
where%
\begin{equation}
\mathcal{J}=\frac{i}{2\pi }\int_{0}^{2\pi }[(\partial _{t}\langle
u_{-}^{k}|)\partial _{k}|u_{-}^{k}\rangle -(\partial _{k}\langle
u_{-}^{k}|)\partial _{t}|u_{-}^{k}\rangle ]\mathrm{d}k
\end{equation}%
is the adiabatic current. Then $c$ is pumped charge of all channel $k$
driven by the time-dependent Hamiltonian varying in a period, which can be
measured through a quasi adiabatic process.

Inspired by these analysis,\ we expect that the Chern number can be unveiled
by the pumping charge of all the energy levels. This can be done in
single-particle sub-space. The accumulated charge passing the unit cell $l$
during the period $T$ is%
\begin{equation}
Q_{l}=\sum_{k}\int_{0}^{T}j_{l}\mathrm{d}t,
\end{equation}%
where current across two neighboring unit cells is%
\begin{eqnarray}
&&j_{l}=\frac{1}{i}\left\langle u_{-}^{k}\left( t\right) \right\vert [\alpha
a_{j}^{\dag }b_{j+1}+\beta b_{j}^{\dag }a_{j+1}+\mu b_{j}^{\dag }a_{j+2}+
\notag \\
&&\nu a_{j}^{\dag }b_{j+2}-i\kappa a_{j}^{\dag }a_{j+3}+i\kappa b_{j}^{\dag
}b_{j+3}-\text{\textrm{H.c.}}]\left\vert u_{-}^{k}\left( t\right)
\right\rangle .
\end{eqnarray}

As we mentioned above, there are three types of adiabatic loop $\mathbf{r=r}%
(t)$ in auxiliary space, with pumping charges $Q_{l}=0$, $1$, and $2$,
respectively. In general, three periodic functions $r_{\bot }(t)$, $\phi (t)$%
, and $z(t)$\ should be taken to measure the pumping charge. However, a
quasi adiabatic loop is tough to be realized in practice. Thanks to the
Biot-Savart law for the field $\mathbf{P}(r)$, we can take the adiabatic
passage along a straight line with fixed $r_{\bot }$ and $\phi $, since the
field $\mathbf{P}$\ far from the trefoil knot $\mathbf{r}_{c}(k)$ has no
contribution to the Amp\`{e}re circulation integral, or the pumping charge.

We consider the case by taking $z=\omega t$ with $\omega \ll 1$. According
to the analysis above, if $t$ varies from $-\infty $ to $\infty $, $Q_{l}$\
should be $0$, $1$, and $2$, respectively. To examine how the scheme works
in practice, we simulate the quasi-adiabatic process by computing the time
evolution numerically for finite system. In principle, for a given initial
eigenstate $\left\vert u_{-}^{k}\left( 0\right) \right\rangle $, the time
evolved state under a Hamiltonian $H_{\text{L}}\left( t\right) $ is%
\begin{equation}
\left\vert \Phi \left( t\right) \right\rangle =\mathcal{T}\{\exp
(-i\int_{0}^{t}H_{\text{L}}\left( t\right) \mathrm{d}t)\left\vert
u_{-}^{k}\left( 0\right) \right\rangle \},
\end{equation}%
where $\mathcal{T}$ is the time-ordered operator. In low speed limit $\omega
\rightarrow 0$, we have%
\begin{equation}
f\left( t\right) =\left\vert \langle u_{-}^{k}\left( t\right) \left\vert
\Phi \left( t\right) \right\rangle \right\vert \rightarrow 1,
\end{equation}%
where $\left\vert u_{-}^{k}\left( t\right) \right\rangle $\ is the
corresponding instantaneous eigenstate of $H_{\text{L}}\left( t\right) $.
The computation is performed by using a uniform mesh in the time
discretization for the time-dependent Hamiltonian $H_{\text{L}}$. In
order to demonstrate a quasi-adiabatic process, we keep $f\left( t\right)
>0.9$\ during the whole process by taking sufficient small $\omega $. Fig. %
\ref{fig4} plots the simulations of particle current and the corresponding
total probability, which shows that the obtained dynamical quantities are in
close agreement with the expected Chern number.

\section{Summary and discussion}

\label{Summary}

We have analyzed a family of $2$D tight-binding model with various Chern
numbers, which are directly connected to the topology of two knots. When
reduced to $1$D single-knot degeneracy model, a polarization vector field
can be established for a gapped band. We have exactly shown an interesting
analogy between the topological feature of the band and classical
electromagnetism: polarization vector field acts as the static magnetic
field generated by the degeneracy knot as a current circuit. It indicates
that there is a quantum analogy of\ Biot-Savart law in quantum matter.
Before ending this paper, we would like to point out that our findings also
reveal the topological feature hidden in the case with zero Chern number. In
Fig. \ref{fig2}(a) and (b), we find out that though the linking numbers of
these two sets of loops are zero, the configurations are different. It
should imply certain topological feature in a single direction, which will
be investigated in future work. This finding extends the understanding of
topological feature in matter and provides methodology and tool for dealing
with the calculation and detection of Chern numbers.

\section{Appendix: Proof of the Biot-Savart law}

In this appendix, we provide the proof of Eq. (\ref{Polarization}) in the
main text. To this end, we first revisit the Biot-Savart law\ for a current
carrying loop, and then compare it with the polarization field in the
present work.

\subsection{The magnetic field}

Consider a current carrying loop $\mathrm{L}$\ with current strength $I=1/%
\mu
_{0}$, which is described by a periodic function $\mathbf{r}_{2}\left(
k_{y}\right) =x_{2}\mathbf{i+}y_{2}\mathbf{j+}z_{2}\mathbf{k}$ in a $3$D
space. Here $%
\mu
_{0}$ is the vacuum permittivity of free space and the current\ flows in the
direction of increasing $k_{y}$ from $0$ to $2\pi $. According to the
Biot-Savart law, the magnetic field $\mathbf{B}$ at position $\mathbf{r}%
_{1}=x_{1}\mathbf{i+}y_{1}\mathbf{j+}z_{1}\mathbf{k}$ generated by the loop $%
\mathrm{L}$ is

\begin{equation}
\mathbf{B}=\frac{1}{4\pi }\oint_{\mathrm{L}}\frac{\mathbf{r}_{2}-\mathbf{r}%
_{1}}{\left\vert \mathbf{r}_{1}-\mathbf{r}_{2}\right\vert ^{3}}\times
\mathrm{d}\mathbf{r}_{2}.
\end{equation}%
For the sake of simplicity we only give the proof for $\mathbf{B}$\ and $%
\mathbf{P}$\ in the $x$ component as an example. The explicit form of the
component is%
\begin{equation}
B_{x}=\frac{1}{4\pi }\oint_{\mathrm{L}}\frac{\left( y_{2}-y_{1}\right)
\mathrm{d}z_{2}\mathbf{-}\left( z_{2}-z_{1}\right) \mathrm{d}y_{2}}{%
\left\vert \mathbf{r}_{1}-\mathbf{r}_{2}\right\vert ^{3}}.
\end{equation}%
According to the Stokes' theorem, the line integral of $B_{x}$ can be
expressed as a double integral%
\begin{eqnarray}
B_{x} &=&\frac{1}{4\pi }\iint\nolimits_{\mathrm{S}}[\frac{3\left(
x_{2}-x_{1}\right) ^{2}-\left\vert \mathbf{r}_{1}-\mathbf{r}_{2}\right\vert
^{2}}{\left\vert \mathbf{r}_{1}-\mathbf{r}_{2}\right\vert ^{5}}\mathrm{d}%
y_{2}\mathrm{d}z_{2}  \notag \\
&&-\frac{3\left( x_{2}-x_{1}\right) \left( y_{1}-y_{2}\right) }{\left\vert
\mathbf{r}_{1}-\mathbf{r}_{2}\right\vert ^{5}}\mathrm{d}z_{2}\mathrm{d}x_{2}
\notag \\
&&-\frac{3\left( x_{2}-x_{1}\right) \left( z_{1}-z_{2}\right) }{\left\vert
\mathbf{r}_{1}-\mathbf{r}_{2}\right\vert ^{5}}\mathrm{d}x_{2}\mathrm{d}%
y_{2}],
\end{eqnarray}
where $S$ represents a smooth surface spanned by the loop $L$.

\subsection{The polarization vector field}

Now we turn to the quantum analogy of Biot-Savart law. For a fixed $k_{x}$, $%
h_{\mathbf{k}}$\ reduces to a $1$D system $h_{k_{y}}$, and the corresponding
Zak phases for upper and lower bands are defined as%
\begin{equation}
\mathcal{Z}_{\pm }=\frac{i}{2\pi }\int_{-\pi }^{\pi }\left\langle u_{\pm
}^{k_{y}}\right\vert \frac{\partial }{\partial k_{y}}\left\vert u_{\pm
}^{k_{y}}\right\rangle \mathrm{d}k_{y},
\end{equation}%
which is gauge-dependent. For the present expression of $\left\vert u_{\pm
}^{k_{y}}\right\rangle $, we have
\begin{equation}
\mathcal{Z}=\mathcal{Z}_{+}=-\mathcal{Z}_{-}=\frac{1}{2\pi }\oint_{\mathrm{L}%
}\cos ^{2}\frac{\theta }{2}\mathrm{d}\varphi ,
\end{equation}%
where $\mathrm{L}$ denotes loop $\mathbf{r}_{2}(k_{y})$\ and%
\begin{equation}
\cos \theta =\frac{z_{1}-z_{2}}{\left\vert \mathbf{r}_{1}-\mathbf{r}%
_{2}\right\vert },\tan \varphi =\frac{y_{1}-y_{2}}{x_{1}-x_{2}}.
\end{equation}%
The polarization vector field is defined as%
\begin{equation}
\mathbf{P}=-\mathbf{\nabla }\mathcal{Z},  \label{P1}
\end{equation}%
where $\mathbf{\nabla }$\ is the nabla operator%
\begin{equation}
\mathbf{\nabla }=(\frac{\partial }{\partial x_{1}}\mathbf{i}+\frac{\partial
}{\partial y_{1}}\mathbf{j}+\frac{\partial }{\partial z_{1}}\mathbf{k}),
\end{equation}%
with unitary vectors $\mathbf{i}$, $\mathbf{j}$, and $\mathbf{k}$ in $3$D
auxiliary space. We note the fact that

\begin{equation}
\mathbf{k\cdot \lbrack }\oint_{\mathrm{L}}\frac{\mathbf{r}_{2\bot }-\mathbf{r%
}_{1\bot }}{\left\vert \mathbf{r}_{2\bot }-\mathbf{r}_{1\bot }\right\vert
^{2}}\times \mathrm{d}\left( \mathbf{r}_{2\bot }-\mathbf{r}_{1\bot }\right)
]=\oint_{\mathrm{L}}\mathrm{d}\varphi =2\pi w,
\end{equation}%
where $w$ is winding number of the integral loop $\mathbf{r}_{2\bot
}(k_{y})=x_{2}\mathbf{i}+y_{2}\mathbf{j}$\ around the point $\mathbf{r}%
_{1\bot }=x_{1}\mathbf{i}+y_{1}\mathbf{j}$. Then the Zak phase can be
rewritten as%
\begin{equation}
\mathcal{Z}=\frac{\mathbf{k}}{4\pi }\mathbf{\cdot }\oint_{\mathrm{L}}\left[
1+\frac{z_{1}-z_{2}}{\left\vert \mathbf{r}_{1}-\mathbf{r}_{2}\right\vert }%
\right] \frac{\mathbf{r}_{2\bot }-\mathbf{r}_{1\bot }}{\left\vert \mathbf{r}%
_{2\bot }-\mathbf{r}_{1\bot }\right\vert ^{2}}\times \mathrm{d}\left(
\mathbf{r}_{2\bot }-\mathbf{r}_{1\bot }\right) .
\end{equation}%
The projection of the polarization vector field $\mathbf{P}$ in the $x$
direction is represented as%
\begin{equation}
P_{x}=-\frac{\partial }{\partial x_{1}}\mathcal{Z}=\frac{1}{4\pi }\oint_{%
\mathrm{L}}\left( G\mathrm{d}x_{2}+Q\mathrm{d}y_{2}+R\mathrm{d}z_{2}\right) ,
\end{equation}%
where%
\begin{eqnarray}
G &=&-\frac{\left( z_{1}-z_{2}\right) \left( x_{2}-x_{1}\right) \left(
y_{2}-y_{1}\right) }{\left\vert \mathbf{r}_{1}-\mathbf{r}_{2}\right\vert
^{3}\left\vert \mathbf{r}_{2\bot }-\mathbf{r}_{1\bot }\right\vert ^{2}} \\
&&-\left( 1+\frac{z_{1}-z_{2}}{\left\vert \mathbf{r}_{1}-\mathbf{r}%
_{2}\right\vert }\right) \frac{2\left( x_{2}-x_{1}\right) (y_{2}-y_{1})}{%
\left\vert \mathbf{r}_{2\bot }-\mathbf{r}_{1\bot }\right\vert ^{4}},  \notag
\end{eqnarray}%
and%
\begin{eqnarray}
Q &=&\left( 1+\frac{z_{1}-z_{2}}{\left\vert \mathbf{r}_{1}-\mathbf{r}%
_{2}\right\vert }\right) \frac{\left[ \left( x_{2}-x_{1}\right) ^{2}-\left(
y_{2}-y_{1}\right) ^{2}\right] }{\left\vert \mathbf{r}_{2\bot }-\mathbf{r}%
_{1\bot }\right\vert ^{4}}  \notag \\
&&+\frac{\left( z_{1}-z_{2}\right) \left( x_{2}-x_{1}\right) ^{2}}{%
\left\vert \mathbf{r}_{1}-\mathbf{r}_{2}\right\vert ^{3}\left\vert \mathbf{r}%
_{2\bot }-\mathbf{r}_{1\bot }\right\vert ^{2}}.
\end{eqnarray}%
By the Stokes' theorem, the line integral of $P_{x}$ can be expressed as a
double integral

\begin{eqnarray}
P_{x} &=&\frac{1}{4\pi }\iint\nolimits_{\mathrm{S}}[\frac{3\left(
x_{2}-x_{1}\right) ^{2}-\left\vert \mathbf{r}_{1}-\mathbf{r}_{2}\right\vert
^{2}}{\left\vert \mathbf{r}_{1}-\mathbf{r}_{2}\right\vert ^{5}}\mathrm{d}%
y_{2}\mathrm{d}z_{2}  \notag \\
&&-\frac{3\left( x_{2}-x_{1}\right) \left( y_{1}-y_{2}\right) \allowbreak }{%
\left\vert r_{2}-r_{1}\right\vert ^{5}}\mathrm{d}z_{2}\mathrm{d}x_{2}  \notag
\\
&&-\frac{3\left( x_{1}-x_{2}\right) \left( z_{2}-z_{1}\right) }{\left\vert
\mathbf{r}_{1}-\mathbf{r}_{2}\right\vert ^{5}}\mathrm{d}x_{2}\mathrm{d}%
y_{2}],
\end{eqnarray}%
which results in%
\begin{equation}
P_{x}=B_{x}.
\end{equation}%
Similarly, the projection of polarization vector field $\mathbf{P}$ and
magnetic field $\mathbf{B}$ in the $y$ and $z$ direction can be calculated
in the same way. Eventually, we can come to a conclusion%
\begin{equation}
\mathbf{P}=\mathbf{B}=\frac{1}{4\pi }\oint_{\mathrm{L}}\frac{\mathbf{r}_{2}-%
\mathbf{r}_{1}}{\left\vert \mathbf{r}_{1}-\mathbf{r}_{2}\right\vert ^{3}}%
\times \mathrm{d}\mathbf{r}_{2}.
\end{equation}

\acknowledgments This work was supported by the National Natural Science
Foundation of China (under Grant No. 11874225)..


\begin{thebibliography}{99}
\bibitem{XWAN} X. Wan, A. M. Turner, A. Vishwanath, and S. Y. Savrasov,
Topological semimetal and Fermi-arc surface states in the electronic
structure of pyrochlore iridates, \textit{Phys. Rev. B} \textbf{83}, 205101
(2011).

\bibitem{LLU} L. Lu, L. Fu, J. D. Joannopoulos, and M. Solja\v{c}i\'{c},
Weyl points and line nodes in gyroid photonic crystals, \textit{Nature
Photon.} \textbf{7}, 294--299 (2013).

\bibitem{SMH} S.-M. Huang, S.-Y. Xu, I. Belopolski, C.-C. Lee, G. Chang, B.
Wang, N. Alidoust, G. Bian, M. Neupane, A. Bansil et al., A Weyl Fermion
semimetal with surface Fermi arcs in the transition metal monopnictide TaAs
class, \textit{Nature Commun}. \textbf{6}, 7373 (2015).

\bibitem{HWENG} H. Weng, C. Fang, Z. Fang, B. A. Bernevig, and X. Dai, Weyl
Semimetal Phase in Noncentrosymmetric Transition-Metal Monophosphides,
\textit{Phys. Rev. X} \textbf{5}, 011029 (2015).

\bibitem{SYXU} S.-Y. Xu, I. Belopolski, N. Alidoust, M. Neupane, G. Bian, C.
Zhang, R. Sankar, G. Chang, Z. Yuan, C.-C. Lee, \ M. Z. Hasan et al.,
Discovery of a Weyl fermion semimetal and topological Fermi arcs, \textit{%
Science} \textbf{349}, 613--617 (2015).

\bibitem{BQLV} B.\thinspace Q. Lv, H.\thinspace M. Weng, B.\thinspace B. Fu,
X.\thinspace P. Wang, H. Miao, J. Ma, P. Richard, X.\thinspace C. Huang,
L.\thinspace X. Zhao, G.\thinspace F. Chen, Z. Fang, X. Dai, T. Qian, and H.
Ding, Experimental Discovery of Weyl Semimetal TaAs, \textit{Phys. Rev. X}
\textbf{5}, 031013 (2015).

\bibitem{LLU2} L. Lu, Z. Wang, D. Ye, L. Ran, L. Fu, J. D. Joannopoulos, and
M. Solja\v{c}i\'{c}, Experimental observation of Weyl points, \textit{Science%
} \textbf{349}, 622--624 (2015).

\bibitem{VMO} V. Mourik, K. Zuo, S. M. Frolov, S. R. Plissard, E. P. A. M.
Bakkers, L. P. Kouwenhoven, Signatures of Majorana Fermions in Hybrid
Superconductor-Semiconductor Nanowire Devices, \textit{Science} \textbf{336}%
, 1003--1007 (2012).

\bibitem{SNA} S. Nadj-Perge, I. K. Drozdov, J. Li, H. Chen, S. Jeon, J. Seo,
A. H. MacDonald, B. A. Bernevig, and A. Yazdani, Observation of Majorana
fermions in ferromagnetic atomic chains on a superconductor, \textit{Science}
\textbf{346}, 602--607 (2014).

\bibitem{PAM} P.A.M. Dirac, Quantized singularities in the electromagnetic
field, \textit{Proc. Royal Soc.} \textit{London} , \textbf{A133} 60 (1931).

\bibitem{DXIAO} D. Xiao, M.-C. Chang, and Q. Niu, Berry phase effects on
electronic properties, \textit{Rev. Mod. Phys.} \textbf{82}, 1959 (2010).

\bibitem{AAB} A. A. Burkov, M. D. Hook, and L. Balents, Topological nodal
semimetals,\ \textit{Phys. Rev. B} \textbf{84}, 235126 (2011).

\bibitem{TB} T. Bzdusek, Q. Wu, A. R\"{u}egg, M. Sigrist, and A. A.
Soluyanov, Nodal-chain metals,\ \textit{Nature} (London) \textbf{538}, 75
(2016).

\bibitem{ZYAN} Z. Yan, R. Bi, H. Shen, L. Lu, S.-C. Zhang, and Z. Wang,
Nodal-link semimetals,\ \textit{Phys. Rev. B} \textbf{96}, 041103 (2017).

\bibitem{RBI} R. Bi, Z. Yan, L. Lu, and Z.Wang, Nodal-knot semimetals,\
\textit{Phys. Rev. B} \textbf{96}, 201305 (2017).

\bibitem{XQSUN} X.-Q. Sun, B. Lian, and S.-C. Zhang, Double helix nodal line
superconductor, \textit{Phys. Rev. Lett.} \textbf{119}, 147001 (2017).

\bibitem{SNIE} S. Nie, H. Weng, and F. B. Prinz, Topological nodal-line
semimetals in ferromagnetic rare-earth-metal monohalides, \textit{Phys. Rev.
B} \textbf{99}, 035125 (2019).

\bibitem{JAHN} J. Ahn, D. Kim, Y. Kim, and B.-J. Yang, Band Topology and
Linking Structure of Nodal Line Semimetals with $Z_{2}$ Monopole Charges,
\textit{Phys. Rev. Lett.} \textbf{121}, 106403 (2018).

\bibitem{CFANG1} C. Fang, H. Weng, X. Dai, and Z. Fang, Topological nodal
line semimetals,\ \textit{Chin. Phys. B} \textbf{25}, 117106 (2016).

\bibitem{TK} T. Kawakami and X. Hu, Symmetry-guaranteed and accidental
nodal-line semimetals in fcc lattice,\ \textit{Phys. Rev. B} \textbf{96},
235307 (2017).

\bibitem{JYL} J. Y. Lin, N. C. Hu, Y. Jian Chen, C. H. Lee, and X. Zhang,
Line nodes, Dirac points, and Lifshitz transition in two-dimensional
nonsymmorphic photonic crystals,\ \textit{Phys. Rev. B} \textbf{96}, 075438
(2017).

\bibitem{CFANG2} C. Fang, Y. Chen, H.-Y. Kee, and L. Fu, Topological nodal
line semimetals with and without spin-orbital coupling,\ \textit{Phys. Rev. B%
} \textbf{92}, 081201 (2015).

\bibitem{PYC} P. Y. Chang and C. H. Yee, Weyl-link semimetals, \textit{Phys.
Rev. B} \textbf{96}, 081114 (2017).

\bibitem{WCHEN} W. Chen, H. Z. Lu, and J. M. Hou, Topological semimetals
with a double-helix nodal link, \textit{Phys. Rev. B} \textbf{96}, 041102
(2017).

\bibitem{MEZAWA} M. Ezawa, Topological semimetals carrying arbitrary Hopf
numbers, \textit{Phys. Rev. B} \textbf{96}, 041202(R) (2017).

\bibitem{YZHOU} Y. Zhou, F. Xiong, X. Wan, and J. An, Hopf-link topological
nodal-loop semimetals, \textit{Phys. Rev. B} \textbf{97}, 155140 (2018).

\bibitem{ZYANG} Z. Yang, C,-K, Chiu, C. Fang, and J. Hu, Jones Polynomial
and Knot Transitions in Hermitian and non-Hermitian Topological Semimetals,
\textit{Phys. Rev. Lett}. \textbf{124}, 186402 (2020).

\bibitem{MXIAO} M. Xiao and S. Fan, Topologically charged nodal surface,\
arXiv:1709.02363.\textbf{\ }

\bibitem{QXU} Q. Xu, R. Yu, Z. Fang, X. Dai, and H. Weng, Topological nodal
line semimetals in the CaP$_{3}$ family of materials, \textit{Phys. Rev. B}
\textbf{95}, 045136 (2017).

\bibitem{RYU} R. Yu, Q. Wu, Z. Fang, and H. Weng, From Nodal Chain Semimetal
to Weyl Semimetal in HfC, \textit{Phys. Rev. Lett.} \textbf{119}, 036401
(2017).

\bibitem{QYAN} Q. Yan, R. Liu, Z. Yan, B. Liu, H. Chen, Z. Wang, and L. Lu,
Experimental discovery of nodal chains, \textit{Nat. Phys.} \textbf{14},
461-464 (2018).

\bibitem{GCHANG} G. Chang, S.-Y. Xu, X. Zhou, S.-M. Huang, B. Singh, B.
Wang, I. Belopolski, J. Yin, S. Zhang, A. Bansil, H. Lin, and M. Z. Hasan,
Topological Hopf and Chain Link Semimetal States and Their Application to Co$%
_{2}$MnGa, \textit{Phys. Rev. Lett.} \textbf{119}, 156401 (2017).

\bibitem{XFENG} X. Feng, C. Yue, Z. Song, Q. Wu, and B. Wen, Topological
Dirac nodal-net fermions in AlB$_{2}$-type TiB$_{2}$ and ZrB$_{2}$, \textit{%
Phy. Rev. Materials }\textbf{2}, 014202 (2018).

\bibitem{RWANG} R. Wang, C. Li, X. Z. Zhang, and Z. Song, Dynamical
bulk-edge correspondence for degeneracy lines in parameter space, \textit{%
Phys. Rev. B} \textbf{98}, 014303 (2018).

\bibitem{XLQI} X.-L. Qi and S.-C. Zhang, Topological insulators and
superconductors, \textit{Rev. Mod.Phys.} \textbf{83}, 1057 (2011).

\bibitem{GYCHO} G. Y. Cho and J. E. Moore, Quantum phase transition and
fractional excitations in a topological insulator thin film with Zeeman and
excitonic masses, \textit{Phys. Rev. B }\textbf{84}\textit{,} 165101 (2011).

\bibitem{RICCA} R. L. Ricca and B. Nipoti, Gauss' linking number revisited,
\textit{J. Knot Theory Ramif.} \textbf{20}, 1325 (2011).

\bibitem{QWZ} X.-L. Qi, Y.-S. Wu, and S.-C. Zhang, Topological quantization
of the spin hall effect in two-dimensional paramagnetic semiconductors,
\textit{Phys. Rev. B} \textbf{74}, 085308 (2006).

\bibitem{SSH} W. P. Su, J. R. Schrieffer, and A. J. Heeger, Solitons in
Polyacetylene, \textit{Phys. Rev. Lett.} \textbf{42,} 1698 (1979).

\bibitem{Zak} J. Zak, Berry's phase for energy bands in solids, \textit{%
Phys. Rev. Lett.} \textbf{62}, 2747 (1989).

\bibitem{Asboth} J. K. Asb\'{o}th, L. Oroszl\'{a}ny, and A. P\'{a}lyi,
\textit{A Short Course on Topological Insulators: Band Structure and Edge
States in One and Two Dimensions}, Lecture Notes in Physics (Springer
International Publishing, Switzerland, 2016).

\bibitem{Clay} R. T. Clay and S. Mazumdar, Cooperative Density Wave and
Giant Spin Gap in the Quarter-Filled Zigzag Electron Ladder, \textit{Phys.
Rev. Lett.} \textbf{94,} 207206 (2005).

\bibitem{Ueda} Y. Shimizu, S. Aoyama, T. Jinno, M. Itoh, and Y. Ueda,
Site-Selective Mott Transition in a Quasi-One-Dimensional Vanadate V$_{6}$O$%
_{13}$, \textit{Phys. Rev. Lett.} \textbf{114,} 166403 (2015).

\bibitem{Jo} T. Zhang and G. B. Jo, One-dimensional sawtooth and zigzag
lattices for ultracold atoms, \textit{Sci. Rep.} \textbf{5,} 16044 (2015).
\end{thebibliography}
\end{document}